# Detection of Dark Web Threats Using Machine Learning and Image Processing

Swetha Medipelly and Nasr Abosata
Dept. of Computer Science and Engineering, Northumbria University,
London, UK

swetha.medipelly@northumbria.ac.uk, nasr.abosata@northumbria.ac.uk

*Abstract*—**This paper aimed to discover the risks associated with the dark web and to detect the threats related to human trafficking using image processing with OpenCV and Python. Apart from that, a development environment was set up by installing TensorFlow, OpenCV and Python. Through exploratory data analysis (EDA), significant insights into the distribution and interactions of dataset features were obtained, which are crucial for evaluating various cyberthreats. The construction and evaluation of logistic regression and support vector machine (SVM) models revealed that the SVM model outperforms logistic regression in accuracy. The paper delves into the intricacies of data preprocessing, EDA, and model development, offering valuable insights into network protection and cyberthreat response.**

## I. INTRODUCTION

The dark web represents the contents posted in darknets and can be accessed only through special software like Tor network, I2P, and Riffle. However, due to its anonymous nature, the dark web is used more for illicit, unlawful, and illegal activities. In the present scenario, this part of the World Wide Web is mainly used for illegal trading of drugs, weapons, forged documents, illegal digital products, cyber-attack services and so on. In 2022, the total darknet market value also increased to 1.5 billion U.S. dollars, which indicate rise in more illegal activities on dark web. In this regard the rest of the paper is organized as follows. Section II provides in-depth information about the different types of dark web crimes; its impacts, challenges faced in detecting the dark web crimes, and suggest the proposed algorithms for detecting the dark web. In section III, this Section provides details about the philosophy, approach, strategy, choice, data collection and analysis method. Section IV section shows the results and findings on how image processing with OpenCV and Python developed a system that examined the pictures associated with human trafficking. The discussion paper reflects on the effective ways of detecting dark web threats in section V. Finlly, section VI pinpoints the key findings of the study and suggests ways for improving future research in similar areas of study.

## II. LITERATURE REVIEW

In the Study Image processing is the editing and analysis of digital images to enhance their visual quality or extract relevant information. TensorFlow and OpenCV are two widely used libraries in the domain of image processing, offering powerful tools and algorithms for a variety of tasks. The paper is based on detection mechanisms for dark web threats. For this reason, first it is important to understand the concept of dark web. This paper provides theoretical information regarding the dark web and its social, economic and ethical impact. Furthermore, this paper focuses on one key issue of dark web, i.e. human trafficking and how image detection algorithm can be sued to deal with this problem. Various algorithms have been discussed in this paper.

### A. DARK WEB

The concept of dark web traces back to other levels of internet, which developed gradually into dark web. The first level is the surface web, which is also termed as open web. It indicates every website which can be accessed easily because the search engines direct them. Alternatively, there are various websites, which are inaccessible as search engines do not direct them to the users, creating deep or invisible web [1]. These websites require being types manually to access them and some of them might require using authentication to access. Weimann [2] stated that deep web is the hidden portion of website, which is not observable to regular users.

### B. DARK WEB Social, Economic and ethical Impacts

Dark web comprises various kinds of financial fraudulent activities. For example, one important kind of fraudulent activity is phishing, where visitors are forwarded to a fake website, which appears like the original website. These fishing sites not only harm the individuals, but also harm the original organizations. However, recognizing these phishing websites is hard because they constantly alter their domains and aesthetic appearances [3]. Therefore, they have no possible fixed structure that bears a resemblance to the standard website. On the other hand, Handalage & Prasanga [4] mentioned that since dark web is related with illegal activities, for example, information breaches and sale of stolen data, typical individuals become victim of such functions, which result in economic loss.

### C. Human trafficking

One important illegal act of dark web is human trafficking. The paper [5] stated that humans are transacted through dark web to perform as sex workers or illegal laborers. These activities are profoundly entrenched in the thought of anonymity given by the dark web, which acts as an entrance for criminal realm. The huge domains accessible in dark web are employed by various groups and organizations, which range from thieves, mafias to terrorists. They employ this platform to communicate, to discuss, to collaborate and to act. They target vulnerable individuals to earn money. Their activities are expanded to international borders. Owing to the untraceable characteristics of illegal activities, law enforcement agencies control them.

*D. Proposed Algorithms*

McGill University and Carnegie Mellon University have developed an algorithm namely '**Info shield**', which is able to recognize resemblances and patterns in web advertisements. There are various limitations of this algorithm because it can recognize every resemblance, which can be misleading or take up too many resources for analyzing criminal cases [6]. Another important algorithm developed by Carnegie Mellon University, namely '**Traffic Jam**', which can examine human trafficking through dark web. This algorithm can also be applied for identifying where the advertisements related with human trafficking are posted. In the year 2019, this AI tool is used in 600 criminal cases on human trafficking [6].

*D. Image recongnition algorithm for detecting darkweb*

Image processing is a way of transforming an image into digital form and to conduct various operations from it. It is a kind of signal dispensation, where pictures are used as images and output is related to the image. Typically, an image processing system comprises managing pictures as two-dimensional signals, while applying already set signal processing method. It is based on three stages. The first stage is importing pictures from scanners. The second stage is evaluating and manipulating the pictures, comprising data compression and picture improvement and spotting patterns which are not to human eyes, such as satellite pictures. The final step is output, where the outcome of the evaluation is presented [7].

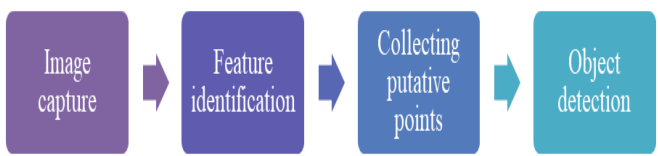

*Fig.1. Object Detection Process*

Fig 1 demonstrates how objects comprising human objects can be detected through image recognition. Finally, a bounding box regression is used to forecast the bounding boxes for every identified region. One of the key fields of AI is the image recognition algorithm, which can recognize objects. Image recognition uses machine vision technologies with AI and trained algorithms to identify pictures by a camera system. Image recognition algorithms work in three stages.

As Shown in the Fig. 2, The first stage is scraping, where web associated with suspected human trafficking is classified. From the websites the pictures are scrapped. The second stage features extraction. In this stage, the pictures are uploaded by the system for detection of trafficking. Various models like HAAR model and algorithms to understand the geometric characteristics of the pictures [8]. Then the third stage is image classification. Through algorithms, the pictures are categorized and accordingly identified by using technologies like Tensor Flow and Open CV. Open CV comprise various computer visualization algorithms, which is used to process the pictures and videos, for recognizing objects and faces. It can be utilized with programming languages like Python, C++, and Java among others. This tool is beneficial for face detection and therefore human traffic observation through dark web [9].

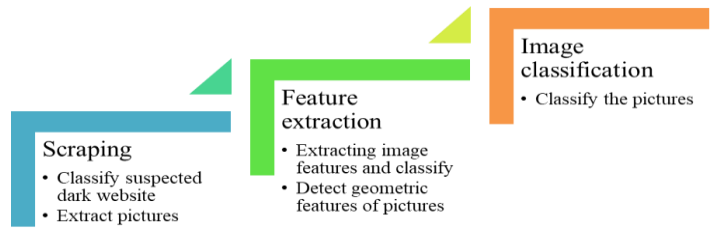

*Fig.2. Image Processing Stages by Algorithms*

## III. METHODOLGY

The aim is to explore the risks associated with the dark web and to detect the threats related to human trafficking using image processing with OpenCV and Python. It thus, important for the researcher to determine appropriate research design and select suitable methodology to achieve this aim of the study and follows positivism philosophy and applies quantitative method based on inductive approach. This study uses experiments as a research strategy. Data in this study is collected from public sources and uses techniques such as image denoising to pre-process the data and remove noise. Next, this study develops Histogram of Oriented Gradients (HOG) algorithm using python for feature extraction and development of Convolutional Neural Network (CNN) for classification. Furthermore, this study uses Accuracy, precision, recall & F1- score for evaluation and optimization of algorithm.

the image processing algorithm such as image loading/acquisition, pre-processing, feature extraction, image classification and detection.

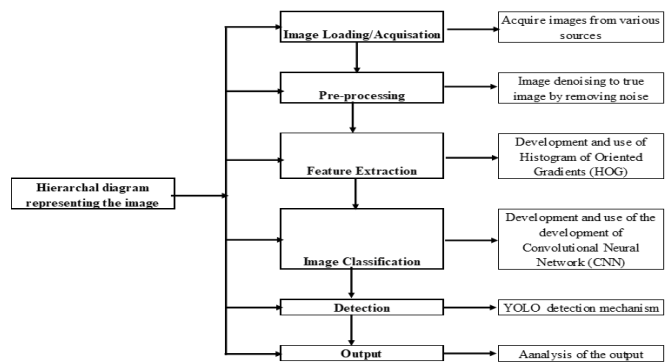

Fig. 3, Hierarchal diagram highlights the key components of

## IV. RESULTS AND FINDINGS

This data comprises cyber security issue reactions, including different kinds of threats as well as the targeted areas. Threats incorporate Malware, Ransomware, Social Designing, and also Phishing, while areas range from Data Breach to Ransomware. Every reaction incorporates the quantity of attempts alongside the overall impact level, alongside an effective target showing whether it addresses the threat of cyber security (1) or not (0). As Shown in the Fig. 4, the dataset offers details into the pervasiveness and effect of various digital threats across areas, helping with the advancement of systems for the location and moderation within cyber security.

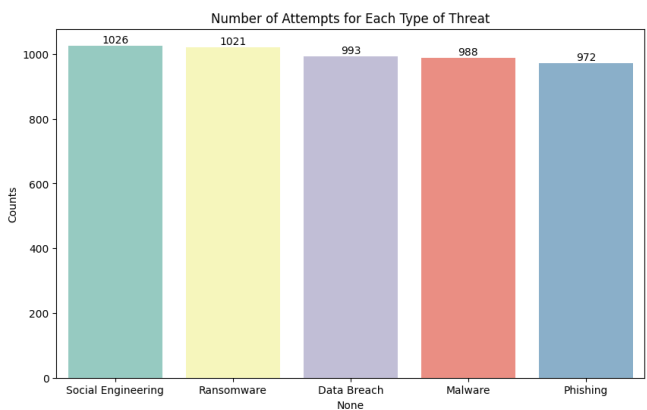

*Fig.4. Showing the pre-processed dataset.*

Fig. 5, Bar diagram delineates the dissemination of the quantity of attempts for every sort of cyber security threat. "Social Engineering" shows the biggest number of attempts, followed intently by Ransomware as well as Phishing. Data Breach alongside the Malware exhibits somewhat lower counts of attempt. The lack of a classification named "None" proposes a total dataset with no missing responses. This representation highlights the predominance and changing powers of various digital threats, giving significant experiences to focusing on threat reaction approaches and allotting resources within cyber security activities.

![Number of Attempts for Each Type of Threat bar chart]

*Fig.5. Number of attempts for every type of threat*

Fig. 6, Histogram portrays the appropriation of the impact levels through digital threat reactions. The specific x-axis addresses the following impact level going from 0 to 100, while the particular y-axis signifies the recurrence of events. The overall distribution shows uniform, with the fluctuations in recurrence seen across various effect levels. This proposes a different scope of effect forces experienced across network protection occurrences. While maximums show moderate effect levels, there are likewise events of higher and lower influence occurrences.

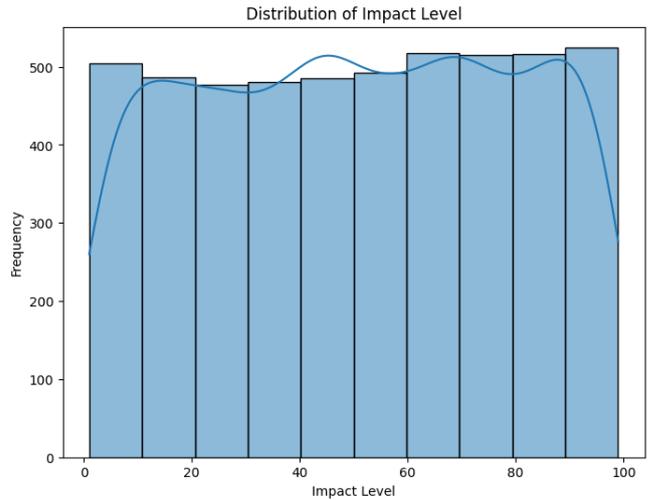

*Fig.6. Distribution of the impact level*

Fig. 7, Distribution of targeted sectors depicts the overall distribution of the targeted areas through digital threats reactions. Every area, for example, Data Breach, Malware and so on, is addressed by a separate-colored section. The respective percentage marks show the extent of events ascribed to every area. The dissemination is somewhat adjusted, with every area representing a comparative portion of designated occurrences. It suggests a far-reaching scope of regions being influenced by network security risks, underlining the unpreventable concept of the respective episodes across various undertakings and progressive spaces. Understanding this dissemination helps with recognizing ordinary targets and shortcomings, enlightening assigned risk balance methods, and protecting endeavors.

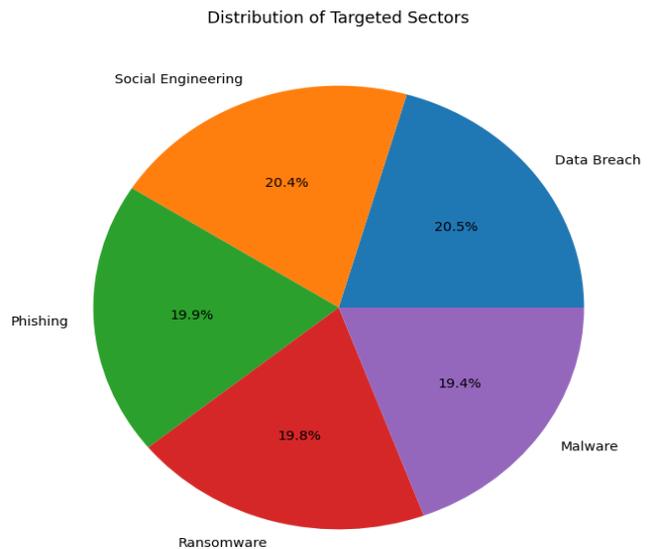

*Fig.7. Distribution of the targeted sectors*

Fig. 8, The Box plot exhibits the dispersal of the number of endeavors for each sort of Organization protection danger. Each part addresses the "interquartile range (IQR)" of the endeavors, with the level line inside showing the middle score. This plot reveals assortments in attempt counts across different threat classifications, with Phishing showing the broadest reach and "Social Engineering" portraying the best median.

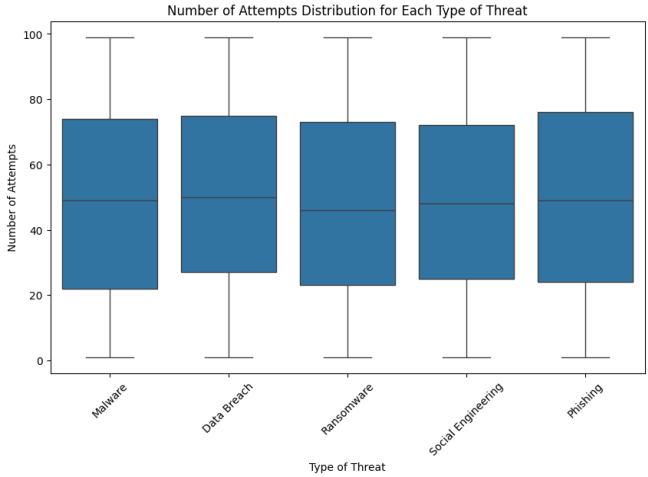

*Fig.8. Number of attempts distribution of every type of threat*

Fig. 9, The heatmap delineates the pairwise connections among various factors within the cyber security data. Every cell within the heatmap addresses the relationship coefficient between two factors, going from - 1 to 1. A score more like 1 demonstrates reliable positive correlation, while a score nearer to - 1 shows major negative correlation. The specific cells show adequate correlations, as they address the connection of every variable with itself. In this heatmap, we notice a moderate positive connection (0.31) between the "Number of Attempts" as well as the "Impact Level," proposing that higher endeavor counts will generally relate with higher effect levels. Moreover, the "Target" variable shows a powerless positive correlation (0.026) with the "Number of Attempts," demonstrating a slight connection between the objective variable and the endeavors made.

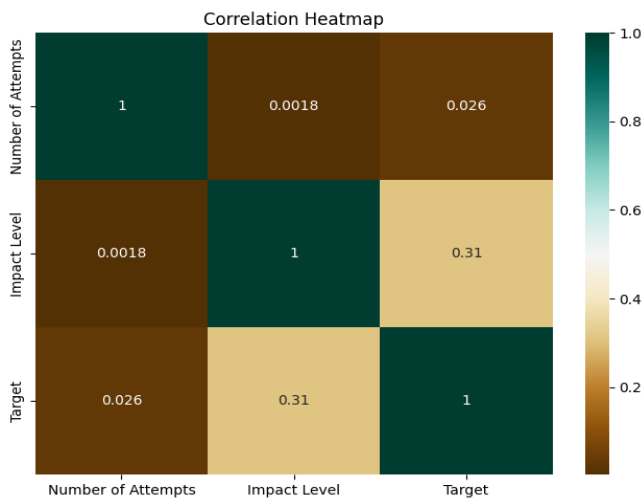

*Fig.9. The Correlation heatmap.*

Fig. 10, addresses an alternate kind of threat, while the blue and orange sections inside each bar portray the quantity of attempts alongside the respective impact level, individually. This representation empowers correlation of the two factors across several threat types. Comparative examples within the quantity of the attempt can be indicated throughout all the threat categories, with phishing displaying eminently greater attempt counts.

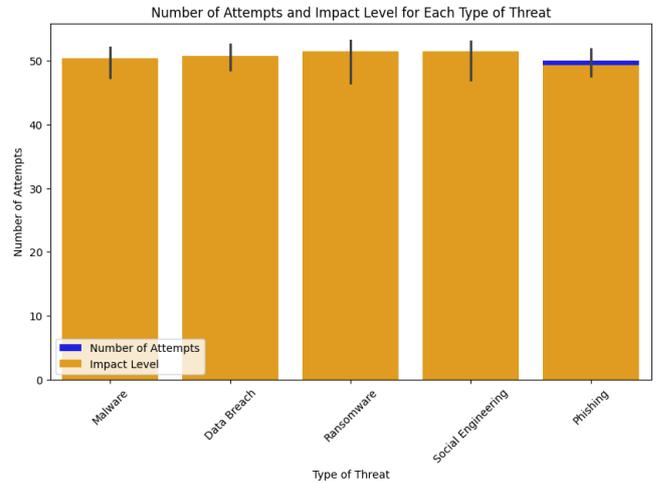

*fig. 10. number of attempts and impact level for every type of threat*

A.

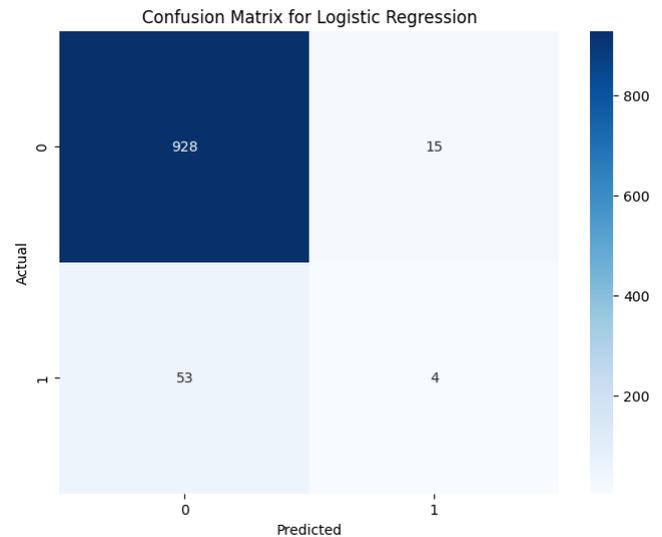

*Fig.11. Confusion Matrix for Logistic Regression*

Fig. 11, confusion matrix is a table that imagines the exhibition of a characterization model by summing up the genuine and anticipated orders. In this confusion matrix: - The upper left cell (928) addresses the quantity of occurrences where the real class was 0 (negative) and the model accurately anticipated it as 0 (true negatives). - The upper right cell (15) demonstrates the quantity of occasions where the genuine class was 0 yet the model erroneously anticipated it as 1 (ales positives). - The base left cell (53) denotes the quantity of cases where the actual class was 1 (positive) yet the model erroneously anticipated it as 0 (False negatives). - There is no sign of true positives in the given matrix. The confusion matrix permits in order to survey the model's exhibition in more detail than precision alone.

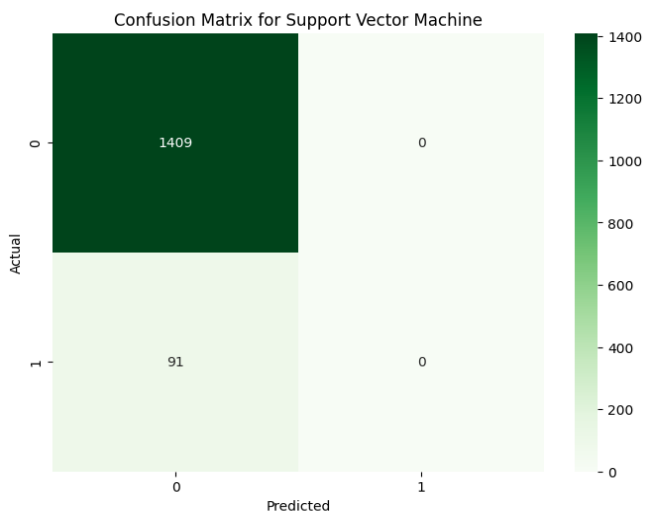

*Fig.12, Confusion Matrix for Support Vector Machine*

Fig. 12, shows the confusion confusion matrix pictures of the exhibition of a support Vector Machine (SVM) model. It comprises four cells addressing various mixes of genuine and anticipated orders. In this disarray lattice: - The upper left cell (1400) demonstrates the quantity of cases where the genuine class was 0 (negative) and the model accurately anticipated it as 0 (true negatives). - The upper right cell (1409) addresses the quantity of occasions where the genuine class was 0 yet the model inaccurately anticipated it as 1 (false positives). - The base left cell (91) implies the quantity of occurrences where the real class was 1 (positive) yet the model erroneously anticipated it as 0 (false negatives). - There is no sign of genuine up-sides in the gave grid. From this framework, it is apparent that the SVM model performs well in anticipating the negative class (0), as demonstrated by the big number of genuine negatives. Be that as it may, it battles with the positive class (1), as proven by the presence of misleading negatives. Generally, the disarray framework gives an unmistakable outline of the SVM model's presentation, featuring its assets and shortcomings in ordering occurrences from various classes. This data is urgent for assessing and refining the model's viability in genuine applications.

Fig. 13, shows the Accuracy Comparison between SVM and Logistic Regression. On the other hand, it can also easily be seen from the above figure that the SVM model got the accuracy as 0.94 and the Logistic Regression model got the accuracy as 0.93. Figure 4.16 portrays a precision examination between Support Vector Machine (SVM) and Logistic Regression models. The correlation shows that the SVM model accomplished an accuracy of 0.94, somewhat beating the Logistic Regression model, which got an accuracy of 0.93. This visual portrayal gives a reasonable correlation of the presentation of these two famous characterization calculations. The higher precision of the SVM model recommends that it very well might be more qualified for the given grouping task contrasted with Calculated Relapse. Notwithstanding, it's fundamental to dig further into different measurements and contemplate past exactness alone to acquire a complete comprehension of model execution. Factors like accuracy, recall, and F1-score can give extra bits of knowledge into the models' capacities to characterize occurrences from various classes accurately. Besides, the decision among SVM and Logistic Regression may likewise rely upon different factors like computational intricacy, interpretability, and the idea of the dataset.

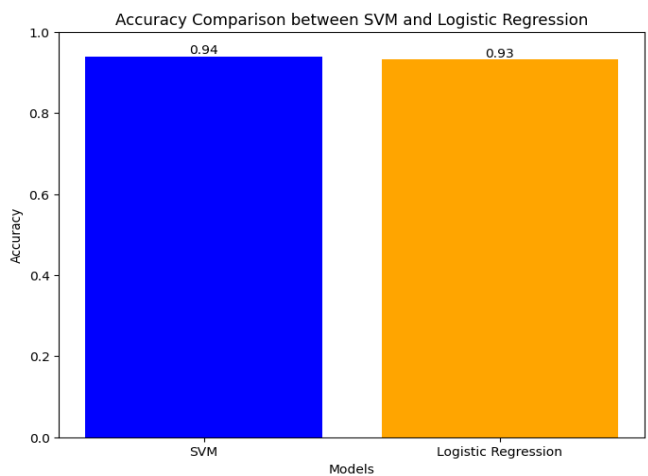

*Fig.13, Accuracy Comparison between SVM and Logistic Regression*

## V. DISCUSSION

From the findings of the study and literature review, it can be discussed that the other significant challenge related to the dark web has been the expansion challenge where online transaction activities tend to be expanded constantly as also the online trading of criminal activities. The main findings associated with the study have been that image recognition algorithms utilizing frameworks such as TensorFlow or Open CV prove to be quite effective in determining photographs that are associated with human trafficking. The findings of the study prove to be important to create and implement proper tools for minimizing human trafficking along with other dark web crimes (Choudhury, 2014). The findings of the study tend to offer major insights into the areas for future research and also offer suggestions to determine dark web criminals and minimize the level of threats. Improving the datasets, enhancing models and also addressing any types of challenges that are faced in detection as well as the location of dark web threats proves to be important for progressing in the field of cybersecurity.

## VI. CONCLUSION

Preprocessing can be considered as the process of preparing data to perform analysis by encoding a wide range of categorical variables, importing the dataset, as well as verifying that no values are null. Hence, understanding the existence as well as the impact of various sorts of distinct cyberthreats can eventually be made possible by the numerous sorts of insights that EDA provides into the distribution as well as interactions among the dataset's many features. In that process of constructing numerous sorts of new models, logistic regression as well as support vector machine models can eventually be used as well as assessed; along with that accuracy comparisons show that the SVM model performs better. The paper dives into the complexities of dataset preprocessing, exploratory data analysis (EDA), and model building, explaining the subtleties of network protection danger responses. Therefore, several recommendations for future studies and strategies in identifying criminals operating on the dark web while minimizing risks were made according to findings from this study.